\documentclass[aps,prb,twocolumn,superscriptaddress,showpacs,floatfix]{revtex4-2}

\usepackage{amsmath,amssymb}
\usepackage{graphicx}
\usepackage{color}
\usepackage{epstopdf}
\usepackage{natbib}
\usepackage{xfrac}
\usepackage{wasysym}
\usepackage{multirow}

\usepackage[colorlinks,breaklinks,bookmarks=true,citecolor=blue,linkcolor=blue,urlcolor=blue]{hyperref}

\usepackage{ulem}

\makeatletter
\let\old@makecaption=\@makecaption
\usepackage{subcaption}

\let\@makecaption=\old@makecaption
\makeatother

\usepackage{xcolor,colortbl}

\definecolor{Gray}{gray}{0.85}
\definecolor{lblue}{rgb}{0.8,0.8,1}  
\definecolor{blue}{rgb}{0.6,0.6,0.9}  

\definecolor{blau}{rgb}{0.,0.,1.}
\definecolor{gruen}{rgb}{0.,0.5,0.0}  \definecolor{orange}{rgb}{1,0.625,0.}  \definecolor{rot}{rgb}{1,0.0,0.}

\newcommand{\fref}[1]{Fig.~\ref{#1}}

\newcommand{\etal}{{\it et al.}}

\renewcommand{\Im}{\hbox{Im}}

\newcommand{\out}[1]{}

\newcommand*{\bo}{\mathbf{0}}

\usepackage{csquotes}

\definecolor{blue}{rgb}{0.,0.,1.}
\definecolor{green}{rgb}{0.,0.5,0.0}  \definecolor{orange}{rgb}{1,0.625,0.}  \definecolor{rot}{rgb}{1,0.0,0.}

\begin{document}
\title{Tuning crystal-fields by He-irradiation and orbital Widom line in SrVO$_3$ films
}

\author{Matthias Pickem}
\affiliation{Institute for Solid State Physics, TU Wien,  A-1040 Vienna, Austria}

\author{Karsten Held}
\affiliation{Institute for Solid State Physics, TU Wien,  A-1040 Vienna, Austria}

\author{Jan M.~Tomczak}
\affiliation{Department of Physics, King's College London, London, United Kingdom}
\affiliation{Institute for Solid State Physics, TU Wien,  A-1040 Vienna, Austria}
\email{jan.tomczak@kcl.ac.uk}

\date{\today}

\begin{abstract}
Helium-ion irradiation of epitaxial SrVO$_3$/SrTiO$_3$ films causes a metal-insulator transition, so far attributed to a Mott localization driven by a reduced kinetic energy. Using density-functional theory plus dynamical mean-field theory, we show that the driving mechanism is instead the crystal-field splitting generated by the irradiation-induced tetragonal expansion, not the interaction-to-bandwidth ratio. The resulting $c$-axis vs.\ temperature phase diagram mirrors that of the one-band Hubbard model, but its critical end-point spawns an {\it orbital} Widom line, rooted in an anomalous compressibility of orbital, rather than charge occupation. At an effectively quarter filling, superexchange-like processes favor orbital over magnetic long-range order. Our results semi-quantitatively reproduce the fluence-dependent spectral gaps and transition thresholds reported experimentally, establishing ion implantation as a route to chemically expand correlated materials.
\end{abstract}

\maketitle

\section{Introduction}

The competition between phases in correlated materials hinges on a complex interplay of various degrees of freedom and their respective energy scales. The canonical Mott transition~\cite{imada,Gebhard1997} requires a sufficiently large ratio of interaction vs.\ kinetic energy to localize electrons. Besides global manipulations, such as hydrostatic pressure~\cite{PhysRevLett.102.237201}, this correlation-strength can be tuned by structural symmetry breaking: the lifting of orbital degeneracy and/or band-width engineering through chemical substitutions~\cite{pavarini_mott_2004,1367-2630-7-1-188,PhysRevB.80.235104}, uniaxial stress~\cite{Eerenstein:2007aa}, substrate strain in film samples~\cite{Shoham2023,jan-strain}, or changes in dimensionality, e.g., quantum-wells in ultrathin films~\cite{yoshimatsu_dimensional_2010,zhong_electronics_2015,GabelP1,pickemZoologySpinOrbital}.

A curious tool to access expanded structures---effectively applying an
inverse pressure otherwise hard to come by---is ion implantation~\cite{Xiang2021,Herklotz2025}. Ideally the ion used in the irradiation should be electronically and magnetically inactive and populate interstitial lattice sites rather than acting as a substitutional defect.
Our motivation stems from the seminal experiment of Wang \etal\ \cite{PhysRevMaterials.3.115001} who applied this technique to the drosophila material of the correlated electron community, SrVO$_3$, and evidenced a metal-insulator transition above a certain irradiation fluence. The situation is enriched by the fact that samples were not bulk SrVO$_3$. For bulk, a simple argument applies: (uniform) chemical pressure leads to an increase of the ratio of the Hubbard  $U$ over the hopping amplitude $t$ and triggers the Mott transition. However, Wang \etal\ studied thin films of SrVO$_3$ grown on an SrTiO$_3$ substrate. This activates, as we shall see, the orbital degrees of freedom. 

In this paper, using  dynamical mean-field theory~\cite{bible,vollkot}
calculations, we will puzzle together a phase diagram that looks from far away like that of the iconic one-band Hubbard model. However, the protagonist in this story is not the Hubbard $U$, but the crystal-field splitting. Correspondingly, the nature of long-range orders and phase boundaries will be intimately linked to the orbital degrees of freedom instead of the charge ones. Among others, we shall discover an {\it orbital Widom line} emerging from the critical end point of the Mott transition, that owes to an anomaly in the compressibility of an orbital's occupation.

\section{Motivation: He-implantation in SrVO$_3$}

\begin{figure}
    \centering
    \includegraphics[trim=45 10 150 20,clip,width=1.02\linewidth]{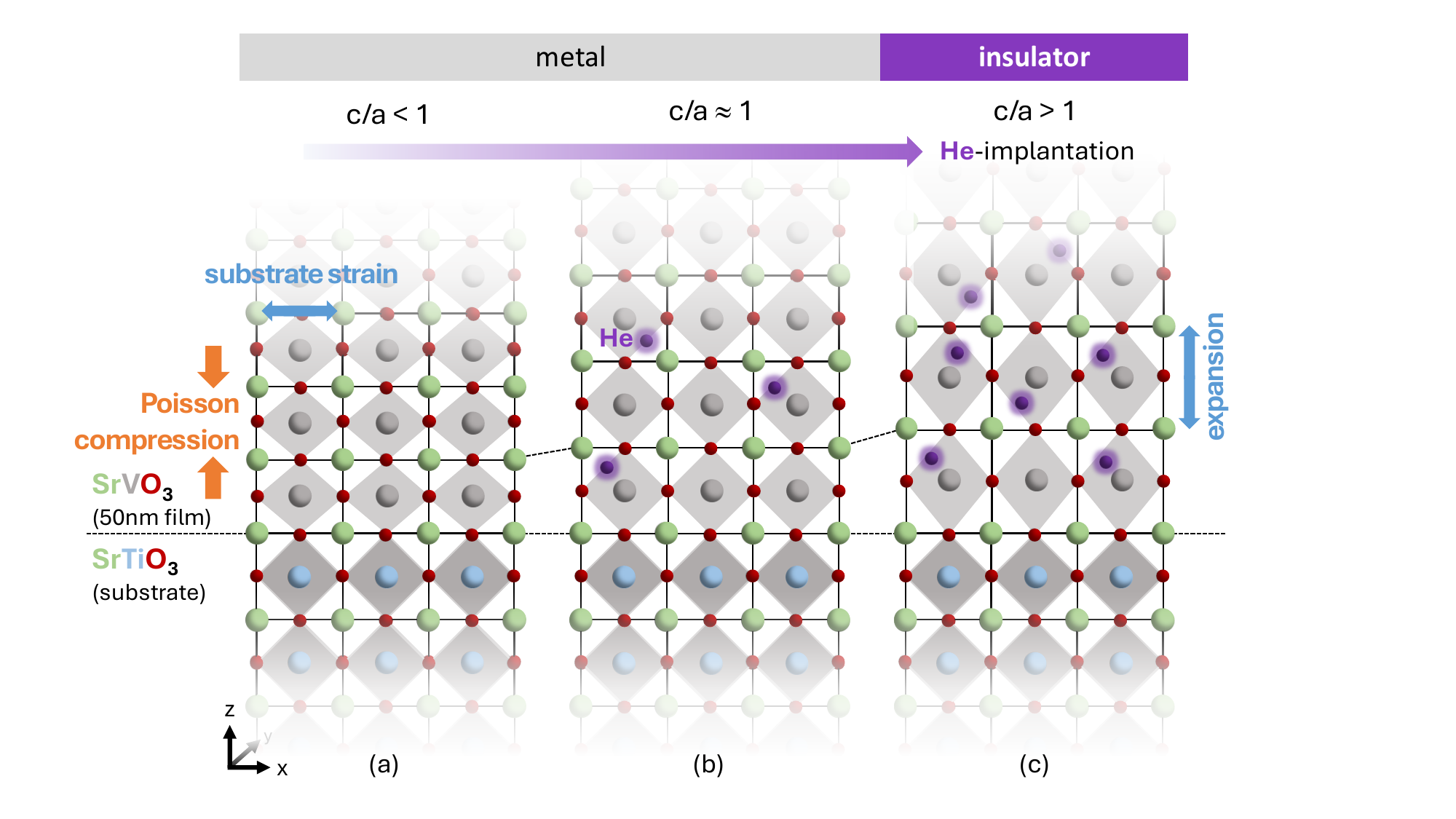} 
    \caption{Schematics of Wang \etal's experiment~\cite{PhysRevMaterials.3.115001}: (a) When grown on SrTiO$_3$, the in-plane lattice constants $a=b$ of SrVO$_3$ are locked by the substrate constraint, leading to a tetragonal compression $c/a<1$. (b) The compression is consecutively overcome by He-ion implantation that acts as chemical pressure; (c) above a critical $c/a>1$ ratio, the system undergoes a metal-insulator transition.}
    \label{fig:scheme}
\end{figure}

Wang \etal~\cite{PhysRevMaterials.3.115001} used helium-ion irradiation to manipulate the out-of-plane lattice constant of $\sim50$nm-thick SrVO$_3$ films grown epitaxially on a single-crystal SrTiO$_3$ substrate, as illustrated in \fref{fig:scheme}:
Perfectly cubic in the bulk ($a=b=c=3.838$\AA~\cite{LandoltBornstein1970:sm_lbs_978-3-540-36202-9_50}), the larger lattice constant of SrTiO$_3$ ($a=b=c=3.988$\AA~\cite{LandoltBornstein1970:sm_lbs_978-3-540-36202-9_50}) causes SrVO$_3$ in films to expand in the $x,y$-plane, while in the $z$-direction the Poisson effect drives a compensatory compression, leading to $c/a<1$; panel (a). 
With the in-plane lattice constant locked to the substrate, implanting Helium atoms results in a uniaxial lattice expansion in the $z$-direction; panels (b) and (c). The inserted He is electronically neutral and X-ray diffraction (XRD) measurements indicate that it does not introduce local crystal defects, suggestive of mediating a near uniform (though not isotropic) chemical pressure~\cite{PhysRevMaterials.3.115001}. Further, the process has been found to be invertible by vacuum annealing~\cite{PhysRevLett.114.256801}, suggesting that the implanted helium atoms populate interstitial sites of the lattice without the creation of atomic vacancies.

Varying the Helium fluence then provides a route to continuously control the tetragonal distortion of the SrVO$_3$ unit-cell
from $c/a<1$ (no or small fluence) to $c/a>1$ (large fluence).
The impact of these structural changes onto the electronic structure were probed by transport measurements:
The pristine sample exhibits metallic (Fermi liquid-like) behavior in the resistivity, $\rho(T)\approx \sigma_0 + bT^2$, 
with $\partial\rho / \partial T > 0$ for all temperatures.%
\footnote{%
Note that recent works~\cite{PhysRevLett.133.186501,467t-z5b2,vjb8-qghf} suggest an active role of electron-phonon effects in the $T^2$-behavior of the resistivity, which is why we shall refer to the resistivity characteristics as Fermi liquid-{\it like}.
}

For a small to intermediate helium fluence ($1-2.5\cdot 10^{15}/$cm$^2$), the resistivity exhibits an overall increase and, at low temperatures, a small upturn emerges, $\partial\rho/\partial T<0$ for $T<T^*$. The onset $T^*$ of deviations to Fermi liquid characteristics moves up in temperature with increasing fluence, from $T^*\approx 40$K to $T^*\approx 70$K for $1.75 \cdot 10^{15}$ to $2.5 \cdot 10^{15}$ He / cm$^2$. 
The positive magneto resistance excludes a dominant weak-localization scenario~\cite{PhysRevMaterials.3.115001} \footnote{%
Irradiation-enabled weak localization (with negative magneto resistance), was found to be the driving force of metal-to-insulator transitions in similar setups for the rare-earth nickelates LaNiO$_3$ and PrNiO$_3$~\cite{PhysRevMaterials.3.053801}.
}.
In congruence with previous work by Fouchet \etal~\cite{FOUCHET20167}, the small resistivity upturn was thus ascribed~\cite{PhysRevMaterials.3.115001}
to interaction corrections to disorder-dressed conduction~\cite{RevModPhys.57.287}%
\footnote{%
Note that Mirjolet \etal~\cite{https://doi.org/10.1002/advs.202004207} suggested that (parts of the) deviations from $T^2$-behavior could also stem from strong electron-phonon coupling effects.
}.

Increasing He-fluence to $3.0 \cdot 10^{15}$ He / cm$^2$ 
triggers another, qualitative change: The resistivity jumps by more than one order of magnitude and exhibits a negative slope,  $\partial\rho / \partial T < 0$, for all measured temperatures---both suggestive of an abrupt metal-to-insulator transition~%
\footnote{%
Similar transitions have been observed in other, related perovskite setups~\cite{lanio3-film,Scheiderer}, too.}.

The goal of this work is to understand this He-implantation driven metal-insulator transition. For 
comprehensiveness, we will not only study uniaxial expansion of the SrVO$_3$ film ($c/a$ increase driven by He-implantation), but also uniaxial compression (decreasing $c/a$).

\section{Methods}

In SrVO$_3$ films thicker than $\gtrsim50$ unit-cell layers, the system can be well approximated by its (strained) bulk geometry~\cite{FOUCHET20167}, as surface effects~\cite{GabelP1} and quantum-well physics~\cite{zhong_electronics_2015,pickemZoologySpinOrbital,PhysRevResearch.4.033253}, important in ultra-thin thin films~\cite{PhysRevLett.104.147601}, can be neglected.
 Fixing the in-plane lattice constant to the theoretical value of bulk SrTiO$_3$ ($a=3.95$\AA), we relax the $c$-axis and internal positions of periodic SrVO$_3$ with density-functional theory (DFT) using the PBE exchange-correlation potential~\cite{PhysRevLett.77.3865} as implemented in \texttt{WIEN2k}~\cite{wien2k2020}. Throughout, we assume that tetrahedral symmetry is maintained and no bond distortions emerge (e.g., through VO$_6$ octahedron rotations). 
Low-energy Hamiltonians are then obtained by constructing $t_{2g}$-derived maximally localized Wannier functions, using \texttt{WIEN2WANNIER}~\cite{wien2wannier} and \texttt{Wannier90}~\cite{wannier90}.

Using the Wannier Hamiltonians, we perform dynamical mean-field theory (DFT+DMFT~\cite{Kotliar2006,doi:10.1080/00018730701619647}) calculations with
\texttt{W2DYNAMICS}~\cite{w2dynamics}, using the Kanamori parametrization of the Hubbard interaction with $U=5$eV, $J_H=0.75$eV and $U'=3.5$eV. These values are typical for DMFT calculations of bulk SrVO$_3$ in a $t_{2g}$-orbitals setting~\cite{pavarini_mott_2004,PhysRevB.73.155112,byczuk-2007-3}.
In principle,
the substrate strain as well as the uni-axial lattice deformation modifies both the size of on-site and exchange terms~\cite{jmtwannier}. Further the lifting of the cubic crystal symmetry will lead to an additional orbital dependence, i.e., $U_{xy,xy,xy,xy} \neq U_{xz,xz,xz,xz} = U_{yz,yz,yz,yz}$, see, e.g., the supplementary material of Ref.~\cite{zhong_electronics_2015}.
Expecting only minor {\it quantitative} differences in the DMFT phase diagram, we forgo these complications here.
We compute susceptibilities using the \texttt{A$\Gamma$GA} code~\cite{CPC_ADGA} and
analytically continue Matsubara Green's functions with the maximum entropy method using \texttt{ana\_cont}~\cite{KAUFMANN2023108519}.

\section{Results}

\subsection{Density Functional Theory -- Tuning the crystal fields}

Relaxing substrate-strained SrVO$_3$ we find a Poisson-ratio driven compression that reduces the $c$-axis lattice constant
from the unstrained bulk value of $a_{\mathrm{SrVO}_3}=c_{\mathrm{SrVO}_3}=3.85$\AA\ to a total-energy optimised value of $c=3.81$\AA, i.e., $c/a<1$, to account for the tensile strain of the substrate.
This value of the $c$-axis lattice constant is what is manipulated, in experiment, through He-implantation. Theoretically, we perform a scan of the $c$-axis parameter, perusing the experimentally realized expansion, while exploring the change of physics under  compression.

\begin{figure}[!t!h]
  \centering
  \includegraphics[trim=4 0 4 0,clip, width=1.025\linewidth]{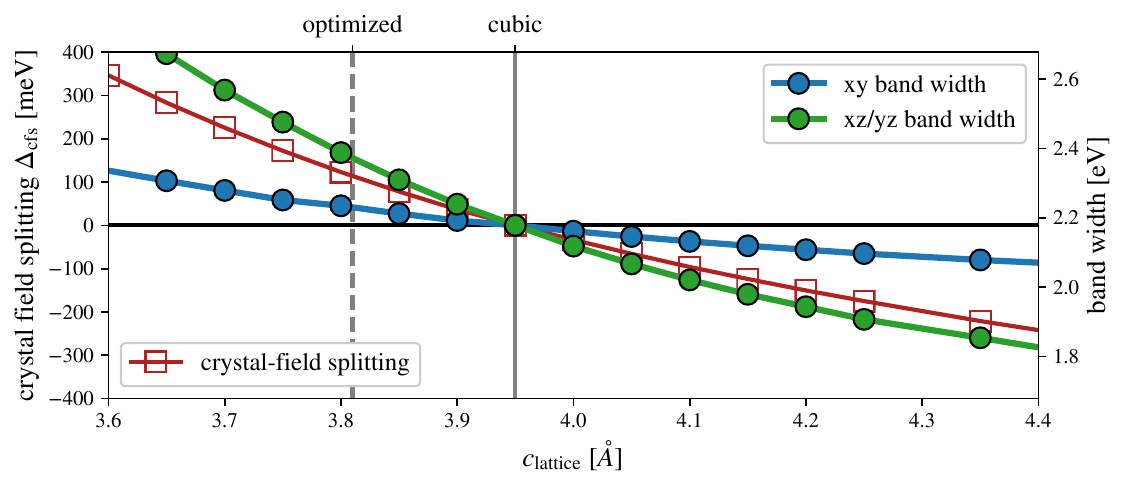} 
  \caption{
  Crystal-field splitting $\Delta_\mathrm{cfs} = E_{xz/yz}-E_{xy}$ (red line, left y-axis) and band widths of the $d_{xy}$ and $d_{xz}$/$d_{yz}$ orbitals (blue and green lines, right y-axis) of tetragonally deformed bulk SrVO$_3$ using DFT. The in-plane lattice constant is constrained to $a=b=3.95$\AA~of the substrate while the perpendicular $c$ axis is varied. For $c<3.95$\AA~the $d_{xy}$-orbital is energetically favored ($\Delta_\mathrm{cfs}>0$), for $c>3.95$\AA~the $d_{xz}$/$d_{yz}$-orbitals are favored ($\Delta_\mathrm{cfs}<0$).}
 \label{fig:srvo3_dft_orbitals}
\end{figure}

A first insight into the electronic structure for varying tetragonal distortion is provided by the one-particle $t_{2g}$-Wannier Hamiltonian. Specifically, we look at the orbital-dependent band widths $W$ and the crystal-field splitting 
\begin{equation}
    \Delta_\mathrm{cfs} = E_{xz/yz}-E_{xy}
\end{equation}
given by the difference of the on-site orbital energies $E_L=\sum_\mathbf{k} H_{LL}(\mathbf{k})$ with $L\in\{\mathrm{xy},\mathrm{xz},\mathrm{yz}\}$ of the Wannier Hamiltonian $H$. 
The result of the $c$-axis scan is displayed in Fig.~\ref{fig:srvo3_dft_orbitals}.
At $a=b=c$ (solid vertical line), cubic symmetry is retained, resulting in local orbital equivalence ($W_{\mathrm{xz}/\mathrm{yz}} = W_\mathrm{xy}$) and a vanishing crystal field splitting $\Delta_\mathrm{cfs} \equiv 0$. 
The bandwidth $W_{a=3.95\text{\AA}} = 2.17\,$eV is smaller compared to $W_{a=3.85\text{\AA}} = 2.42\,$eV of bulk SrVO$_3$, 
owing to a decreased overlap of orbitals at larger volume. 
Energetically optimizing the $c$-parameter naturally breaks this symmetry establishing a tetragonal lattice with a compressed out-of-plane lattice of $c=3.81$\AA~causing a finite crystal-field splitting of $\Delta_\mathrm{cfs} = +120$meV to develop (see dashed vertical line in Fig.~\ref{fig:srvo3_dft_orbitals}).
In this equilibrium, the in-plane tensile straining is compensated by an increased out-of-plane $d_{xz}$/$d_{yz}$ orbital overlap, increasing their intra-band hopping amplitudes. 
This distortion then results in a joint (though uneven) band width increase of both the $d_{xy}$ and $d_{xz}$/$d_{yz}$ orbital to $W_\mathrm{xy}=2.22$eV and $W_{\mathrm{xz}/\mathrm{yz}} = 2.36$eV, respectively.
As expected, the $d_{xz}$/$d_{yz}$ band width is more sensitive to $c$-axis manipulation than the $d_{xy}$ band width.
On the opposite end, the effects are reversed: Expanding the out-of-plane axis against its equilibrium tendency results in a positive crystal field splitting, now energetically favoring the $d_{xz}$/$d_{yz}$ orbitals.
By further elongating the vanadium-vanadium bonds away from equilibrium, both band widths decrease in value.
The band-structures at representative $c/a$ ratios can be found in Fig.~\ref{fig:srvo3_dft_bscomp} of the Appendix.

\subsection{Dynamical mean-field Theory}

\subsubsection{Stabilizing the Mott insulator}

While Wang \etal~\cite{PhysRevMaterials.3.115001} performed DFT+DMFT calculations and attributed the irradiated samples' tendency to turn insulating to electronic correlations, they did not actually find insulating solutions in their simulations.
We now provide a possible perspective for the hitherto elusive, correlation-driven metal-insulator transition responsible for the observed trends seen in the transport measurements.

\begin{figure*}[!t!h]
  \centering
  \begin{subfigure}{0.75\linewidth}
  \raggedright (a) 
  \includegraphics[width=\linewidth]{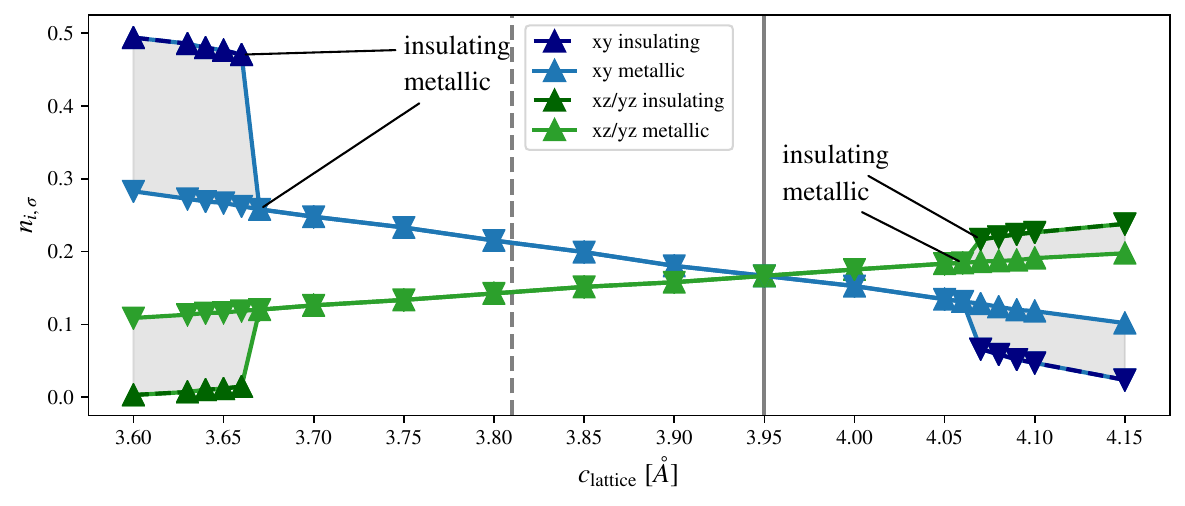}
  \phantomsubcaption\label{fig:srvo3_dmft_occ}
  \end{subfigure}
\vspace{-5mm}
\begin{subfigure}{0.75\linewidth}
    \raggedright (b) $\qquad\qquad\qquad\qquad\qquad\quad\quad \quad\qquad \,$
    $\qquad\qquad\quad\quad$ (c) 
    \includegraphics[width=\linewidth]{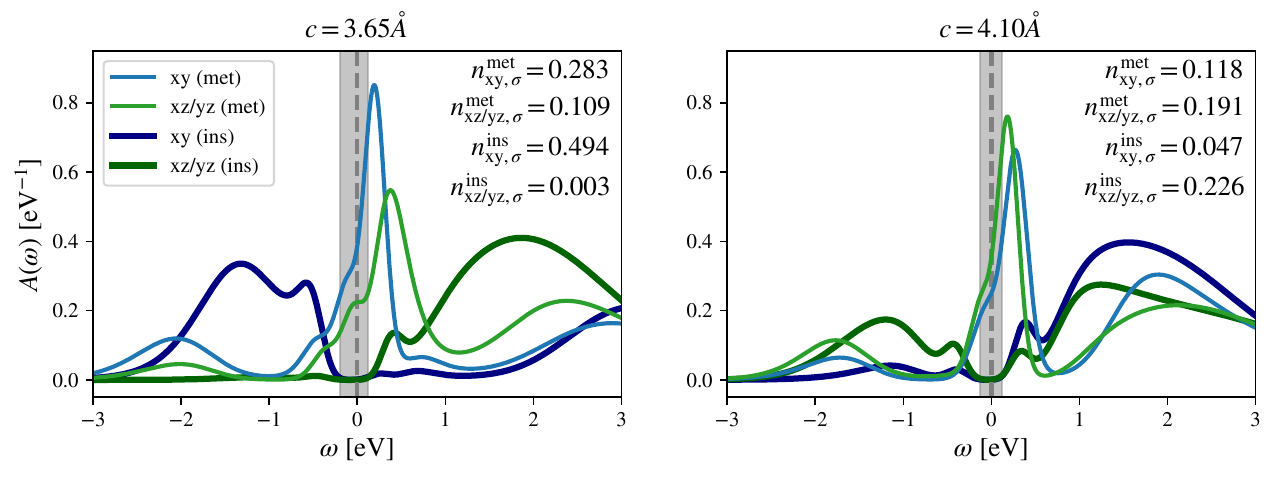}
     \phantomsubcaption\label{fig:srvo3_dmft_spectral}
\end{subfigure}
\vspace{-5mm}
\begin{subfigure}{0.75\linewidth}
    \raggedright (d) $\qquad\qquad\qquad\qquad\qquad\quad\quad  \quad\qquad \,$ 
    $\qquad\qquad\quad\quad$ (e) 
    \includegraphics[width=\linewidth]{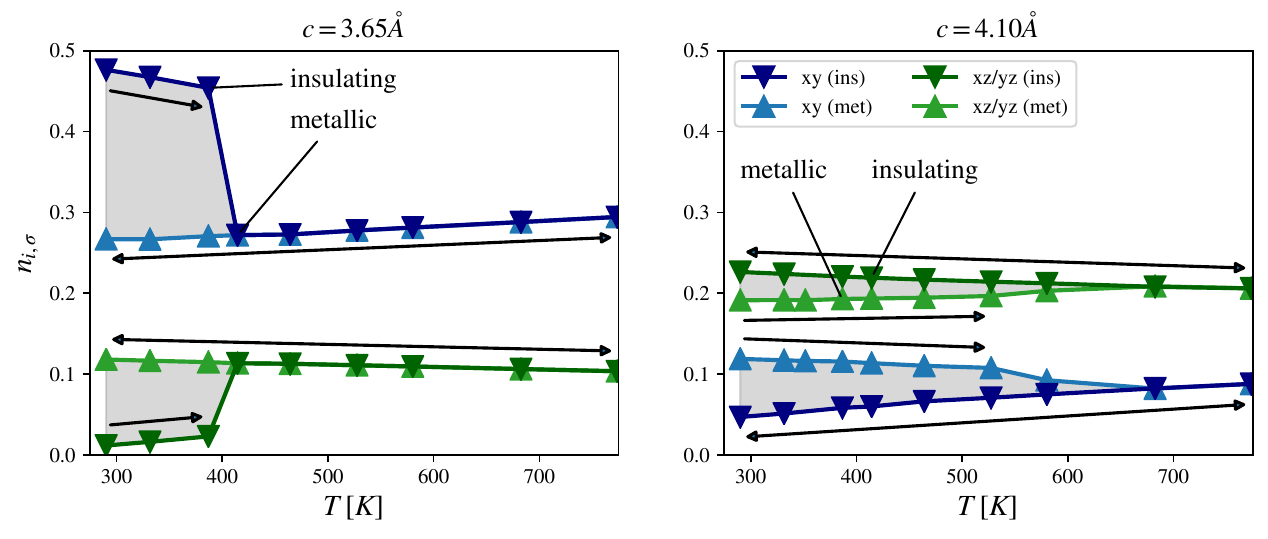}
    \phantomsubcaption\label{fig:srvo3_temp}
\end{subfigure}
  \caption{
  Dynamical mean-field theory (DMFT) results.
  (a) Orbital and spin resolved occupations at room temperature ($T=290\,$K, $\beta=40\,$eV$^{-1}$). The sudden jump in orbital polarization for small and large $c$-axis parameters signals polarized insulating solutions. 
  (b,c) 
  Orbital resolved local spectral functions in the coexistence region of uniaxially compressed ($c=3.65$\AA; panel b) 
  and elongated ($c=4.10$\AA; panel c) bulk SrVO$_3$ with an in-plane lock to $a=b=3.95$\AA, at  room temperature, $T=290$K.
  The insulating solutions, with spectral gaps marked in gray, are characterized by a nearly half-filled $d_{xy}$ orbital [$n_{\mathrm{xy},\sigma} \approx 0.5$; panel (b)] and a quarter-filled $d_{xz}$/$d_{yz}$ doublet [$n_{\mathrm{xz}/\mathrm{yz},\sigma} \approx 0.25$; panel (c)], respectively.
  (d,e)  Orbital occupations of the $c$-axis compressed 
  [$c=3.65$\AA, panel (d)] and expanded structure [$c=4.10$\AA, panel (e)] as a function of temperature, following paths indicated by black arrows.
  Note that cooling stabilizes a low-temperature metallic solution for $c=3.65$\AA\ (panel d) and an insulating solution for $c=4.10$\AA.
  }
\end{figure*}

First, we perform a DMFT $c$-axis scan at fixed room temperature $T=290$K 
and illustrate the resulting orbital occupations in Fig.~\ref{fig:srvo3_dmft_occ}.
Crucially, we are able to stabilize a metallic solution for all considered $c$-axis values, when initializing the simulations with  a {\it metallic starting point} (e.g., $\Sigma = 0$).
In line with the crystal-field splitting, the system naturally prefers occupying the lower-lying energy level(s) resulting in an orbital polarization mirroring the crystal-fields from \fref{fig:srvo3_dft_orbitals}.
Moving away from the cubic $c=3.95$\AA~($\forall L: n_{L,\sigma}=1 / 6$), we find an almost perfect linear variation of the occupations with $c$-axis strain.
We theorize that this linear dependence results from the interplay of the crystal-field splitting and the band widths:
The faster-than-linear increase (decrease) of $\Delta_\mathrm{cfs}$ in the compressed (expanded) regime, is counterbalanced with a simultaneous increase (decrease) of orbital band widths, see Fig.~\ref{fig:srvo3_dft_orbitals}.

Once metallicity is established in the system, we find that $c$-axis manipulation alone is unable to trigger a transition into an insulating solution for the shown range, i.e.\ for the given temperature we are effectively locked onto the linear occupation branch.
This observation might be the reason why the previously performed calculations~\cite{PhysRevMaterials.3.115001} were only able to find this metallic branch.
However, it turns out that, at the largest compression or elongation of the $c$-axis shown, the system is actually inside the coexistence region (gray shaded area) of a first order Mott metal-insulator transition. That is, the system supports an insulating solution (indicated in Fig.~\ref{fig:srvo3_dmft_occ} with darker colors), but---within the realistic change of the $c$-axis lattice parameter considered---we do not reach the asymmetry required to close the hysteresis loop.  

In order to overpower the metallic solution we resorted to a technical ``trick'', based on the inevitability of a Mott insulating state in integer-filled systems above a finite critical interaction~\cite{PhysRevB.55.R4855}:
We find that, $U\geq 6$eV ($U'=4.5$eV) is sufficient to trigger the Mott insulator for $c\leq 3.85$\AA\ and $c\geq 4.00$\AA, irrespective of the DMFT starting point.
The artificial increase in interaction parameters is then slowly relaxed to the standard values.
Within a finite $c$-range (marked in shaded grey), the system does not revert back to the occupations of the metallic solution.
Instead, a state with a much increased orbital polarization persists.

In the compressed structure ($c \leq 3.66$\AA) we find a state closely resembling the one-band Hubbard model: The $d_{xy}$-orbital is almost half-filled, $n_{xy,\sigma}\lessapprox 0.5$, while the out-of-plane orbitals $d_{xz}$, $d_{yz}$ are essentially empty.
In the expanded structure ($c \geq 4.07$\AA) we find the system to resemble an ideal two-orbital, quarter-filled Hubbard model with $n_{xz,\sigma}=n_{yz,\sigma}\lessapprox 0.25$.

For two representative $c$-axis parameters,  $c = 3.65$\AA\ and $c = 4.10$\AA, 
the resulting orbital-resolved spectral functions are reported in Fig.~3(b) and (c), respectively, and confirm the above picture: under compression, the system exhibits a Mott gap for the half-filled $d_{xy}$-orbital, while the $d_{xz}$/$d_{yz}$ orbitals are unoccupied. The elongated system realizes a Mott gap at quarter filling of the $d_{xz}$/$d_{yz}$ doublet, with
the third orbital being only scarcely filled.
The spectral gaps of $\Delta_\mathrm{DMFT} \approx 350$meV and $\Delta_\mathrm{DMFT} \approx 250$meV in panel (b) and (c), respectively, are uncharacteristically small for Mott insulators.
Indeed, one naively expects gaps of the order $\Delta_\mathrm{DMFT}\sim U$.
The reason for this is that the charge gap in both cases is between states or different orbital characters, while the separation of the Hubbard bands of the same orbital is indeed of the scale $U$.
Similar orbital effects occur in LaTiO$_3$ (Ti with nominal $d^1$ filling) where due to the inter-orbital fluctuations at the small crystal-field (3 t$_\mathrm{2g}$ $\rightarrow$ a$_\mathrm{1g}$ + 2 degenerate e$_\mathrm{g}$ through a GdFeO$_3$ distortion), the observed band gap remains quite small: $\Delta=0.2$eV~\cite{PhysRevB.89.161109}. The same is true for many other multi-orbital materials, including the textbook example for the Mott transition, V$_2$O$_3$~\cite{keller:205116}.

\subsubsection{Temperature dependence}
To further analyze the DMFT solutions we now study the effect of temperature.
In Fig.~3 (d) and (e), we report the orbital occupations as a function of temperature 
for $c=3.65$\AA\ and $c=4.10$\AA, respectively. Starting from $T=290$K, for which the corresponding spectra are shown in panel (b) and (c), we follow both the metallic and the insulating solution to higher temperatures.

In the compressed structure ($c=3.65$\AA, panel (d)),
the insulating solution is found to be stable up to $T=390$K, above which the orbital polarization can no longer be maintained.
Increased thermal fluctuations abruptly collapse the orbital polarization, destroying the insulating solution.
Similarly to the previous $c$-axis scan, once this metallic solution is established, the system remains ``locked'' onto the metallic temperature branch.
This is confirmed by starting from a high-temperature, metallic solution and cooling the system down into the coexistence region (see arrows in Fig.~3(d)).
Once the system is metallic, further temperature increase then only leads to a minor softening of the three-peak structure in the spectral function (see the total spectral functions in \fref{fig:srvo3_dmft_tempbranch1} below), and a slight variation of the orbital occupation.
Cooling on the metallic branch leads to a reduction in orbital polarization $n_{\mathrm{xy},\sigma}-n_{\mathrm{xz/yz},\sigma}$.
This is in contrast to the insulating branch, where, if the polarized Mott insulator is established, cooling further pushes the system towards stronger orbital polarization, eventually yielding a  half-filled $d_{xy}$-orbital: $\lim_{T\rightarrow 0} n_{\mathrm{xy},\sigma}(T) = 0.5$.

The experimentally relevant expanded structure ($c=4.10$\AA) exhibits the opposite
temperature behavior to the compressed structure.
Both metallic and insulating solutions remain intact up to $T\approx 570$K.
At higher $T$, the two branches recombine to a single
{\it insulating} solution with residual weight around the Fermi level, see the spectra in \fref{fig:srvo3_dmft_tempbranch2} below, indicative of a so-called bad insulator.
Contrary to the compressed system, here the metallic branch is the odd one out:
heating the system beyond the critical threshold temperature ($T\approx 650$K), the quasi-particle peak suddenly collapses, its weight being distributed to the two satellite Hubbard bands. 

The linear behavior in the orbital occupation, previously observed for the metal, now occurs for the insulating branch instead.
We verified that cooling from this high temperature solution maintains the bad insulator,
eventually stabilizing the spectral gap of the Mott insulator (see Fig.~\ref{fig:srvo3_cooling}; right panel).
Akin to the polarized insulator found for $c=3.65$\AA, the orbital polarization eventually yields a complete depopulation of the lower lying orbitals for $T\rightarrow 0$: The nominally one electron is equally distributed among the quarter-filled $d_{xz}$, $d_{yz}$ orbitals with $\lim_{T\rightarrow 0} n_{\mathrm{xz/yz},\sigma}(T) = 0.25$, see Fig.~3(e).

\subsubsection{Orbital Widom line}
\begin{figure}[!htb]
  \centering
  \includegraphics[trim=4 0 4 0, clip=true, width=1.02\linewidth]{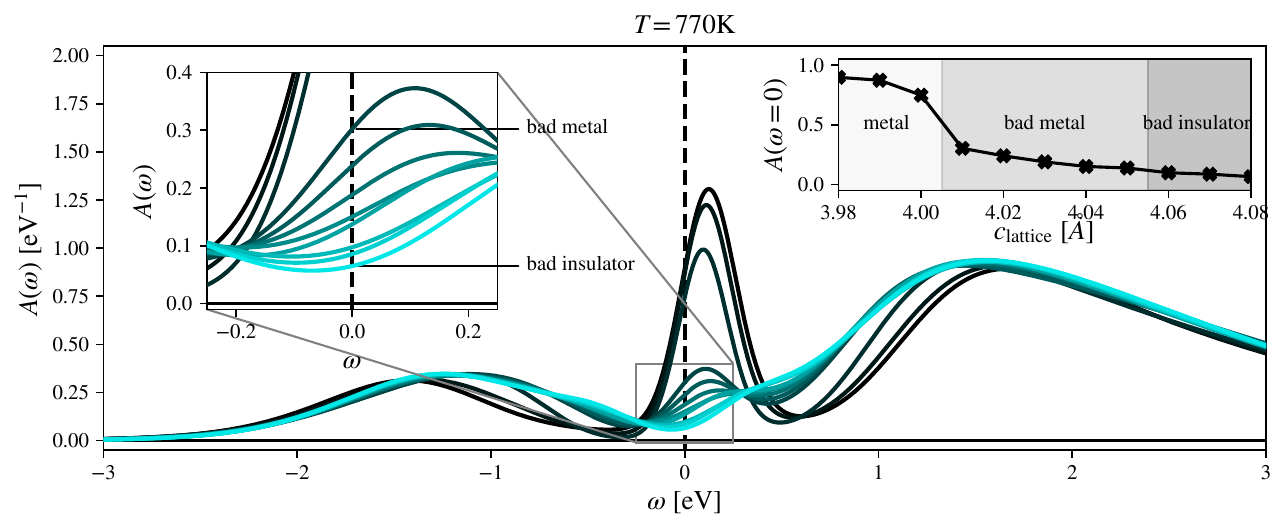}
  \caption{
  Detailed evolution of the spectral function across the metal-insulator crossover as a function of the $c_{\mathrm{lattice}}$ parameter
    (at the 11 values given by the inset)
   at $T=770$K. Beyond a crystal-field splitting threshold, the metal morphs into a badly metal phase (quasi-particle peak is suppressed but remains finite and convex in $\omega$). Beyond $c\gtrsim 4.06$\AA, the spectral function is indicative of a bad insulator (spectral weight at the Fermi level is finite but concave in $\omega$).}
 \label{fig:srvo3_badmetal1}
\end{figure}

We now analyze the $c$-axis dependence of the expanded structure at large temperatures in more detail.
Fig.~\ref{fig:srvo3_badmetal1} displays the spectral function at $T=770$K through the metal-insulator crossover realized by consecutive small increments of the $c$-axis lattice parameter.
The (Fermi liquid-like) metal, with its prominent and convex quasi-particle peak is stable up to $c=4.00$\AA. There the system enters a bad metal regime, characterized by the existence of a finite but largely suppressed quasi particle peak.
Increasing the $c$-axis parameter further, the local maximum of the quasi-particle peak slowly morphs into a local minimum (see the left inset in Fig.~\ref{fig:srvo3_badmetal1}): Beyond around $c=4.06$\AA, the system is a bad insulator, where
the quasi particle peak has vanished but spectral weight nonetheless remains finite $A(\omega=0)>0$. 

\begin{figure*}[!htb]
  \centering
  \includegraphics[width=0.8\linewidth]{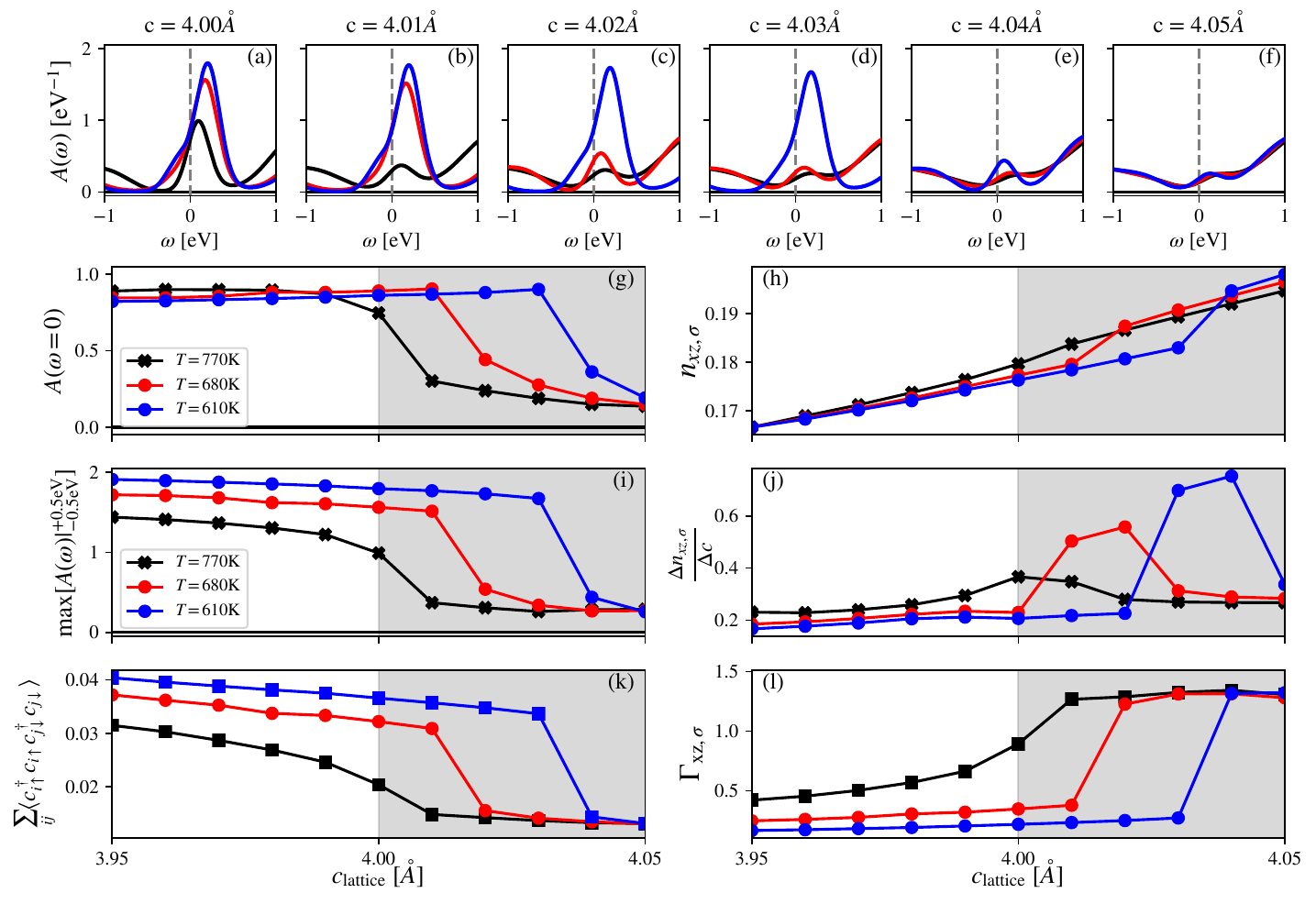}
  \caption{(a-f) Evolution of the spectral function along the metal to bad metal/insulator crossover from $c=4.00 \rightarrow 4.05$\AA. We show the crossover for the temperatures $T=770$K, $T=680$K and $T=610$K. (g) Spectral weight at the  Fermi level $A(\omega=0$) and (i) height  of the quasi particle peak $\max[A(\omega)\vert_{-0.5\mathrm{eV}}^{+0.5\mathrm{eV}}]$. Cooling the system boosts the quasi-particle peak and a larger crystal-field splitting (larger $c$ values) is necessary to turn the system insulating. The crossover is characterized by a sudden change in orbital occupation $n_{\mathrm{xy},\sigma}$ (panel h), and the resulting maximum of the (symmetric) numerical derivative thereof (panel j). Further, the metal-insulator crossover is accompanied by a drop in the total number of double occupancies (panel k) and an enhancement of the electronic scattering rate (panel l).
   \label{fig:srvo3_badmetal}
}
\end{figure*}

We now analyze the temperature dependence of the bad-metal to bad-insulator crossover above the critical point using the results displayed in \fref{fig:srvo3_badmetal}.
Cooling the sample, we find the crystal-field splitting-driven transition to become sharper as a function of the lattice parameter, see panels (a)-(f), where besides the previous $T=770$K (black) we also display results for $T=680$K (red) and $T=610$K (blue).

Further to the full spectra, we show the spectral function evaluated at the Fermi level $A(\omega=0)$ (panel g), the height of the quasi-particle peak (panel i), the $d_{xz}$-orbital occupation (panel h), its derivative with respect to the $c$-axis  (panel j), the total number of double occupancies 
$\sum_{LL'} \left\langle n_{L\uparrow} n_{L'\downarrow}\right\rangle$ 
($L, L' \in \{\mathrm{xy},\mathrm{xz},\mathrm{yz} \}$) (panel k), as well as the scattering rate $\Gamma=-\Im\Sigma(i\nu_n\rightarrow 0^+)$ (panel l).

At constant $c$, cooling leads to an enhancement of the weight (area) of the quasi-particle peak
(although $A(\omega=0)$ is hardly affected within the metallic phase).
Through this stronger metallicity, the system is more resilient against lattice deformation: a larger crystal-field splitting is necessary to trigger the insulator.
The change in quasi-particle weight is reflected in the orbital occupations, see panel (h).
To enhance the visibility of the charge redistribution, we compute derivatives of the occupation using finite differences.
Already at $T=770$K the derivative $\partial n_{xz}/\partial c\approx \Delta n_{xz}/\Delta c$, see panel (j), reveals a regime change at $c=4.0$\AA. When lowering temperature, the crossover becomes more apparent, while moving to larger $c$.
Inspired by the Widom line in the one-band Hubbard model ($\max_\mu 1/n^2 \partial n / \partial\mu$), see, e.g., Ref.~\cite{sordi-widom}, we are able to characterize the metal-to-insulator crossover above the critical point via
\begin{equation}
    c^\textrm{oWL}(T): \max_c \left[\frac{\partial n_{\mathrm{xz}/\mathrm{yz},\sigma}(T)}{\partial c} \right]
\label{eq:orbwidom}
\end{equation}
which we coin ``orbital Widom line'' (oWL).
We theorize that this line originates from the critical end-point of the Mott metal-insulator transition, see Fig.~\ref{fig:srvo3_sketch} for a semi-quantitative sketch.
Moving towards the critical point via cooling, the peak in $\frac{
\partial n_{\mathrm{xz}/\mathrm{yz},\sigma}}{\partial c}(T)$ quickly grows until it will eventually diverge at the end-point.
At and below the critical point the metal-insulator transition is then characterized by the emergence of a discontinuous orbital polarization, see Fig.~\ref{fig:srvo3_dmft_occ}, upholding a firm insulating solution%
\footnote{An alternative connection to the ``ordinary'' Widom line is in the signature of the charge channel: The characteristic electronic compressibility $\kappa_e = 1/n^2 \partial n / \partial\mu$ can be connected to the charge response $\chi_D(\mathbf{Q}=\bo,i\omega_m=0)$, showing a maximum in the doped one-band Hubbard model~\cite{PhysRevLett.125.196403}.
In our setting, crossing the orbital Widom line, we find a maximum of the corresponding {\it eigenvalue} of the density channel $\lambda_D(\mathbf{Q}=\bo,i\omega_m=0)$ which is however, surprisingly, not reflected in any significant changes of the physical response $\chi_D(\mathbf{Q}=\bo,i\omega_m=0)$ (not shown).}.
Across the orbital Widom line, we, unsurprisingly, also find a drop-off in the total number of double occupancies $\sum_{LL'}\langle n_{L\uparrow}n_{L'\downarrow}\rangle$, as well as a strong enhancement of the electronic scattering rate $\Gamma = -i\Im\Sigma(i\nu_n\rightarrow 0^+)$ in the $d_{xz}$/$d_{yz}$ orbitals. For the latter we employed a third order polynomial fit on the first four data points on the Matsubara axis and plot the extrapolated value at $i\nu_n \rightarrow 0^+$ in Fig.~\ref{fig:srvo3_badmetal}(l).
Upon cooling, the drop in double occupancies becomes sharper and larger in size, where below the (anticipated) critical point the continuous transition must eventually transform into a kink, mirroring the discontinuous emergence of the orbital polarization.

\subsubsection{Ordering instabilities}
So far, we discussed DMFT solutions with an enforced SU$(2)$ symmetry, i.e., we ignored any ordering instabilities.
However, depending on the lattice geometry, much of the metal-insulator transition of the one-band Hubbard model is masked by an antiferromagnetic phase. While bulk SrVO$_3$ is paramagnetic throughout, we previously evidenced a "zoo" of phases with long range order in ultra-thin SrVO$_3$ films~\cite{pickemZoologySpinOrbital}. 
The commonality between ultrathin films and the thicker films considered in this work is that 
the $d_{xy}$ vs.\ $d_{xz}$/$d_{yz}$ crystal-field splitting is the qualitative driver of the physics.
In ultrathin films the {\it sign} of the crystal-field splitting changed with the surface termination of the film. Here, instead, we can also tune the {\it magnitude} of the crystal-field splitting through ion implantation, but the argument is similar:

In the compressed structure, say, at $c=3.65$\AA, the crystal-field $\Delta_{\mathrm{cfs}}>0$ causes orbital occupations in the insulating branch to realize a situation similar to the one-($d_{xy}$)-orbital Hubbard model at strong-coupling. 
There, the Heisenberg superexchange $J=-t^2 / U$ can lead to processes that lower the overall energy of the system 
if adjacent spins become anti-aligned: antiferromagnetism emerges.
Given a site $i$ and its neighbor $i+1$, instead of a uniform occupation $(n_{\uparrow},n_{\downarrow})$, i.e.,
\begin{equation*}
    (0.5,0.5)_{i-1} \rightarrow (0.5,0.5)_{i} \rightarrow (0.5,0.5)_{i+1} \rightarrow \ldots
\end{equation*}
the system will favor a alternating configuration
\begin{equation*}
    (1.0,0.0)_{i-1} \rightarrow (0.0,1.0)_{i} \rightarrow (1.0,0.0)_{i+1} \rightarrow \ldots
\end{equation*}
that breaks spin symmetry, leading to, e.g., checker-board magnetic order.
As the hypercubic lattice provides ideal conditions for such alignments, non-local fluctuations begin to proliferate and compete with thermal fluctuations, eventually leading to a second-order magnetic phase transition at finite temperature.
Away from these idealized (half-filled, cubic) conditions, order may still emerge
with a reduced transition temperature. Further, a phase-transition dome is formed as a function of a pertinent control-parameter, such as orbital filling.
The $d_{xy}$-orbital in the multi-orbital structure at $c=3.65$\AA\ represents such a case. The orbital polarization pushes this orbital into the vicinity of half-filling, where we expect antiferromagnetic fluctuations to form.

We quantify these qualitative statements by computing DMFT susceptibilities$\chi$.
While the metallic solution remains firmly non-magnetic (the largest magnetic response found is $\chi_M(\mathbf{Q}=(\pi,\pi,0)) = 10$eV$^{-1}$), the insulating solution appears to be 
below the critical antiferromagnetic ordering temperature at room temperature, $T=290$K. This can be seen by inspecting the leading eigenvalues of the magnetic Bethe-Salpeter equation which are found at $\mathbf{Q}=(\pi,\pi,0)$ and $\mathbf{Q}=(\pi,\pi,\pi)$ with $\lambda^M_{(\pi,\pi,0)} = 1.62$ and $\lambda^M_{(\pi,\pi,\pi)} = 1.68$, respectively. Note that ordering sets in when $\lambda$ passes $1$ from below, i.e.,  above room temperature where we  already have $\lambda>1$.
This  signals that the massive orbital polarization is accompanied by an antiferromagnetic  phase transition.%
\footnote{Please note that we cannot discern which real-space spin arrangement is preferable as we cannot approach the phase transition from higher temperatures. In fact, beyond $T=390$K, the insulating solution breaks down and only the metallic solution (without any sign of ordering) remains, see Fig.~\ref{fig:srvo3_temp}.}

Similarly, in the expanded structure ($c=4.10$\AA) the crystal field $\Delta_{\mathrm{cfs}}<0$ pushes the system towards an effective quarter-filled two-orbital setting, that is prone to so-called orbital ordering. 
Indeed, due to the tetragonal symmetry the low-lying $d_{xz}$, $d_{yz}$ orbitals retain their local degeneracy, providing ideal conditions for order~\cite{juelich2016}.
Here, the ordering stems, assuming only intra-orbital hopping, from virtual hopping processes where the energy gain of roughly $\Delta E\propto -t^2 / U'$ is now inversely proportional to the inter-orbital repulsion $U'<U$.
This ordering tendency can already be observed on the metallic branch: despite being relatively far away from the ideal condition, we find an eigenvalue of the density Bethe-Salpeter equation of $\lambda^D_{(\pi,\pi,\pi)}=0.92$ at room temperature $T=290$K.
Pushing the system insulating redistributes the orbital occupation sufficiently to get close to $n_{xz/yz,\sigma}=0.25$.
There, we evidence commensurate ordering through the eigenvalue in the density channel $\lambda^D_{(\pi,\pi,\pi)}=1.5>1$. That is, the system favors an alternating occupation of $d_{xz}$ and $d_{yz}$ orbitals throughout the lattice.
Instead of a uniform $(n_{xz,\sigma},n_{yz,\sigma})$
\begin{equation*}
    (0.25,0.25)_{i-1} \rightarrow (0.25,0.25)_{i} \rightarrow (0.25,0.25)_{i+1} \rightarrow \ldots
\end{equation*}
the system will realize
\begin{equation*}
    (0.5,0.0)_{i-1} \rightarrow (0.0,0.5)_i \rightarrow (0.5,0.0)_{i+1} \rightarrow \ldots
\end{equation*}
where $i$, again, corresponds to any lattice position $\mathbf{R}_i$ and the alternation is generated along all three directions, corresponding to $\mathbf{Q} = (\pi,\pi,\pi)$ %
\footnote{
Contrary to the manually suppressed antiferromagnetic  order (via spin symmetrization after each DMFT iteration), signatures of the orbital structure can already be seen on the one-particle level in DMFT:
For low enough temperatures and/or large enough (negative) crystal-field splitting (e.g., $c=4.45$\AA, $T=290$K: not shown)
DMFT convergence fails without more elaborate mixing schemes.
The orbital occupations (and self-energy) try to break the (local) symmetry of the Hamiltonian, split asymmetrically and alternate from iteration $n$ to iteration $n+1$, mimicking the order the system wants to, but cannot develop. 
Strictly speaking, for this reason we were not able to continue the $c$-axis scan of Fig.~\ref{fig:srvo3_dmft_occ} to larger values and probe the other side of the indicated hysteresis. There, an explicit symmetry enforcement of the $d_{xz}$/$d_{yz}$ orbitals or a supercell setup would be necessary to converge the DMFT calculation into its orbitally degenerate or orbitally ordered state, respectively.
}.

\section{Discussion}

\subsubsection{Tetragonal dichotomy}
The metal-insulator transitions found at the extremities of the $c$-axis scan revealed a noticeable asymmetry: 
On the expanded side (realized in experiment), a  Mott insulator could be stabilized by cooling the system from high to low temperatures. The $c$-axis compressed structure lacks an insulating phase at high-temperatures, i.e., for small $c$ the metal-insulator coexistence region has an upper critical temperature. This is why, for compressed structures, we had to resort to manipulating the Hubbard interaction to stabilize the insulator in the coexistence region.

Let us highlight this asymmetry, or dichotomy in the tetragonal deformation further:
The insulating solution in the compressed coexistence region at room temperature requires a crystal-field splitting that is roughly three times larger than its expanded counterpart: $\Delta_\mathrm{cfs}(c=3.65\mathrm{\AA})=+120$meV, $\Delta_\mathrm{cfs}(c=4.10\text{\AA})=-40$meV.
From the point of view of orbital degeneracies one would actually expect the effective half-filled one-orbital system realized under stark compression to more readily cross the Mott metal-insulator transition than the effective quarter-filled two-orbital model under $c$-axis expansion. Indeed, in a degenerate $N$ orbital system with $1$ electron per site the critical interaction strengths scales like $U_c(N) \propto \sqrt{N}\,U_c(N=1)$~\cite{PhysRevB.55.R4855}. This naive argument, however, neglects the impact of kinetic changes:
The tendency to remain metallic under $c$-axis Poisson compression is driven by the band-width increase of the $d_{xz}$/$d_{yz}$ orbitals.
Not only does the increased hopping lower the correlation strength (``$U/t$''), but the larger bandwidth also makes it 
harder to depopulate these orbitals, as is required to distil the effective one-band Hubbard model physics.

\subsubsection{Phase diagram and validation to experiment}
\label{sec:sketch}
\begin{figure}[!t!h]
  \centering
  \includegraphics[trim={0 0.5cm 0 -0.5cm}, width=1.02\linewidth]{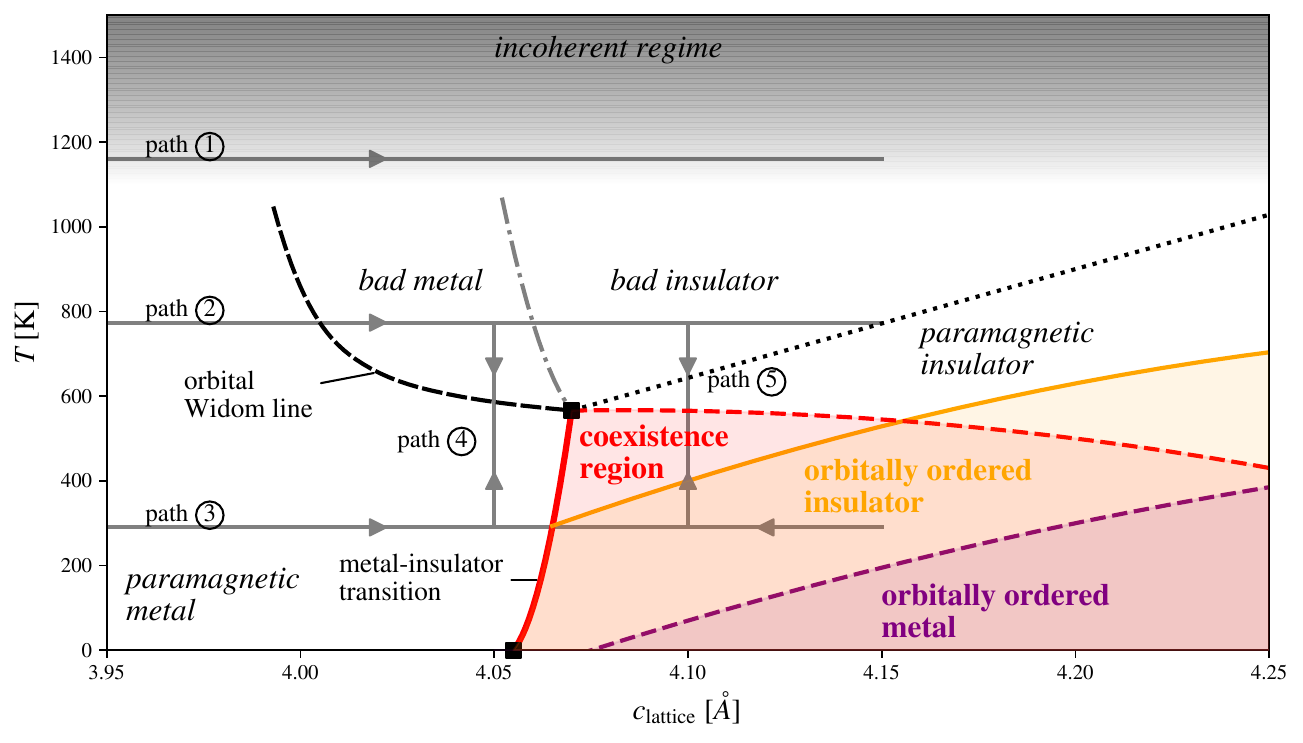}
  \caption{
  Semi-quantitative phase diagram of the expanded structure $c>a=b=3.95$\AA. At elevated temperatures such as $T=1160$K, the system is incoherent. 
  Along the high-temperature path \raisebox{.5pt}{\textcircled{\raisebox{-.9pt} {1}}}, the system smoothly crosses over from a metallic into a bad insulator solution, see Fig.~\ref{fig:srvo3_hightemp2}.
  At lower temperatures, $T=770$K (path \raisebox{.5pt}{\textcircled{\raisebox{-.9pt} {2}}}: Fig.~\ref{fig:srvo3_badmetal1}),
  the crossover sharpens and we can distinguish three phases: The metal is separated from the bad metal by the orbital Widom line; the bad metal can be qualitatively distinguished from the bad insulator by the curvature of the spectral function. 
  At $T=290$K (path \raisebox{.5pt}{\textcircled{\raisebox{-.9pt} {3}}}, the system is below the critical end-point of the crystal-field driven Mott transition (black square). We thus find a sharp metal-insulator transition (red, solid), accompanied by a wide coexistence region (red), cf., Fig.~\ref{fig:srvo3_dmft_occ}).
  Depending on the value of the $c$-axis (or the crystal-field splitting) we can enter the coherent metallic (path \raisebox{.5pt}{\textcircled{\raisebox{-.9pt} {4}}}: Fig.~\ref{fig:srvo3_cooling}) and the coherent insulating (path \raisebox{.5pt}{\textcircled{\raisebox{-.9pt} {5}}}: Fig.~\ref{fig:srvo3_temp} and Fig.~\ref{fig:srvo3_cooling}) regime via cooling.
  In both the insulating and metallic regime, owing to the vicinity of ideal quarter-filling, orbital order eventually emerges (orange and purple dome, respectively), i.e., a spontaneous breaking of the degeneracy of the $d_{xz}$/$d_{yz}$ orbitals.
  Due to the enhanced orbital polarization, the ordering of the insulator necessarily occurs at higher temperatures.}
 \label{fig:srvo3_sketch}
\end{figure}

For the purpose of comparing our result to the experiment of Wang \etal~\cite{PhysRevMaterials.3.115001}, we now restrict ourselves to the analysis of the expanded structure.
We begin by combining the gained insight from our DMFT calculations into a complex temperature vs.\ $c$ lattice constant phase diagram, shown in Fig.~\ref{fig:srvo3_sketch}.
There, we indicate the parameter-paths discussed above with numbered gray lines and indicate the phase boundaries we can infer from the physics along them. Note that the emerging phase diagram is largely schematic as we have quantitative information only along a few parameter trajectories that are marked as path {\textcircled{\raisebox{-.9pt} {1}}} to {\textcircled{\raisebox{-.9pt} {5}}} in Fig.~\ref{fig:srvo3_sketch}.

At large temperatures, thermal fluctuations suppress the formation of any quasi-particle peak for all lattice constants $c\geq 3.95$\AA. This is confirmed by the spectra shown in Fig.~\ref{fig:srvo3_hightemp2} (below) for a $c$-axis sweep at $T=1160$K, marked as ``path \raisebox{.5pt}{\textcircled{\raisebox{-.9pt} {1}}}''. Note that this temperature is not yet sufficient enough to fully destabilize the metallic solution for $c<3.95$\AA, where a modest quasi-particle peak remains.
Cooling the system to intermediate temperatures $T=770$K, see Fig.~\ref{fig:srvo3_badmetal1}, 
gives rise to the first detectable, distinctive regime caused by the crystal-field splitting: path \raisebox{.5pt}{\textcircled{\raisebox{-.9pt} {2}}}. Below a critical $c=4.00$\AA~a coherent, paramagnetic metal is realized, displaying the characteristic three-peak structure. Beyond this $c$-axis threshold (the orbital Widom line at this temperature; black dashed), a different phase emerges: the quasi-particle peak collapses giving way to, first, a bad metal, and, beyond, to a bad insulator solution.

\begin{figure*}[!t!h]
  \centering
  \includegraphics[width=0.8\linewidth]{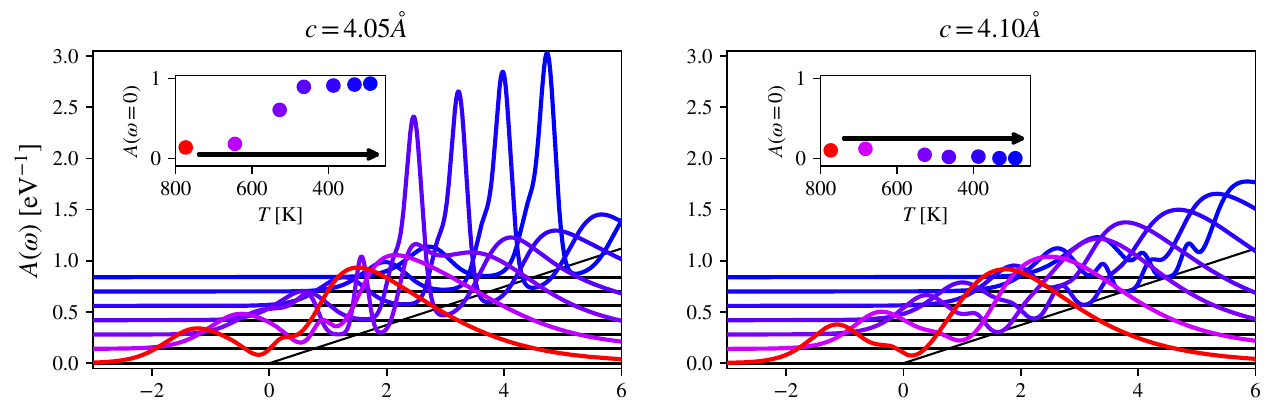}
  \caption{Spectral function evolution for $c=4.05$\AA~(left) and $c=4.10$\AA~(right) upon cooling (the temperatures can be read off from the data point with the same color in the inset). Separated by the critical line, the former enters the coherent metallic regime while the latter enters the insulating regime. These two paths correspond to the trajectories \raisebox{.5pt}{\textcircled{\raisebox{-.9pt} {4}}} and path \raisebox{.5pt}{\textcircled{\raisebox{-.9pt} {5}}} in Fig.~\ref{fig:srvo3_sketch} and explain the metal-insulator transition observed by Wang \etal~\cite{PhysRevMaterials.3.115001} as a crystal-field splitting-induced phenomenon.
  }
 \label{fig:srvo3_cooling}
\end{figure*}

Cooling further, we anticipate the orbital Widom line to end at a critical end-point (black square), below which
we find the opening of a wide metal-insulator coexistence region. There, the metallic solution is valid up to surprisingly large $c$-values: path \raisebox{.5pt}{\textcircled{\raisebox{-.9pt} {3}}}.
We find two distinct transitions from the established bad metal/insulator regions:
Along the vertical path \raisebox{.5pt}{\textcircled{\raisebox{-.9pt} {4}}} ($c=4.05$\AA, $T=770$K $\rightarrow$ $T=290$K) the system changes from bad to coherent, paramagnetic metal whereas at larger $c$, path \raisebox{.5pt}{\textcircled{\raisebox{-.9pt} {5}}} ($c=4.10$\AA, $T=770$K $\rightarrow$ $T=290$K) illustrates the entrance into the coherent, insulating regime, cf.~Fig.~\ref{fig:srvo3_temp} and Fig.~\ref{fig:srvo3_cooling}, respectively.

Once this coherent insulating solution has been established, cooling further naturally stabilizes the accompanying orbital polarization~
\footnote{In order to further characterize the stability of this behavior, a free energy comparison of the two coexisting solution, similar to Ref.~\cite{PhysRevLett.115.256402}, would be in order. At this moment, we do not have access to this expression within our employed impurity solver.}.
In turn, the slope of the transition line (solid red) between the paramagnetic metal and the coexistence region must be positive. Eventually, in the insulating state, orbital ordering will emerge, breaking the local degeneracy of the $d_{xz}$/$d_{yz}$ orbitals.
The shape of the orbitally ordered dome (orange) is also well founded: larger $c$-values result in larger $\Delta_\mathrm{cfs}$. The increased orbital polarization then pushes the system closer to the ``ideal'' quarter-filling, thus increasing the transition temperature.
Please note that we have also theorized that in the metallic regime of the coexistence region (bounded from above by the dashed red line) we also expect orbital order to set in eventually ($\lambda_D=0.92$ for $c=4.10$\AA~and $T=290$K).
Due to the reduced orbital polarization in the metal, the transition temperature (dashed purple line) must necessarily be smaller compared to the insulator.
Please note that it is {\it a priori} not clear whether the orbital-ordering and the metal-insulator-transition boundaries are separated, coincide or overlap.

Having established the theoretical phase diagram, we now connect to and rationalize the behavior seen in Wang \etal's experiment~\cite{PhysRevMaterials.3.115001}.
Samples were epitaxially grown at elevated temperatures of $\approx 920$K, i.e., far above our anticipated critical end-point of the Mott transition.
Cooling down any {\it compressed} structure in the considered range ($3.60$\AA$<c<3.95$\AA), e.g., the non-irradiated thin films, will yield a Fermi liquid solution at room temperature. Indeed, the system is metallic at high-$T$ and cooling follows the metallic-branch of the $c$-axis hysteresis-loop, as discussed above.
The situation for structures sufficiently expanded by He-implantation is different:
Beyond a critical $c$-axis value (the orbital Widom line at that temperature), the high-$T$ solution is {\it insulating}, see the  $c$-axis scan at $T=770$K in Fig.~\ref{fig:srvo3_badmetal1}. When cooling down from there, the system then remains on the insulating branch of the $c$-axis hysteresis-loop (provided that the temperature path does not cross the Widom line).

Quantitatively, the $c$-axis scan at room temperature, path \raisebox{.5pt}{\textcircled{\raisebox{-.9pt} {3}}}, reveals a hysteresis onset which is compatible with the $3.0\cdot10^{15}$ ($c_\mathrm{exp} = 4.03$\AA) and $3.5\cdot10^{15}$ He / cm$^2$ fluence samples ($c_\mathrm{exp} = 4.05$\AA) of Ref.~\cite{PhysRevMaterials.3.115001}.
The minor quantitative difference is likely to 
stem from the DFT setup: Within PBE the lattice constant of SrTiO$_3$ is $a_\mathrm{STO}^\mathrm{PBE} = 3.95$\AA. This is a slight overestimation compared to the experimental value of $a_\mathrm{STO}^\mathrm{EXP} = 3.905$\AA. In order to generate the same critical crystal-field splitting in the real sample, the expanded sample would need a smaller crystal elongation compared to our calculations. This is evident in the smaller experimental $c$-values necessary for the onset of an activated behavior in the transport data.

In the two samples with the largest applied fluence, a negative slope, $\partial\rho/\partial T<0$, is observed for $T<300$K~\cite{PhysRevMaterials.3.115001}, with thermal activation behavior, $\rho(T) \propto \exp\left({\frac{\Delta}{2 k_B T}}\right)$ below $T\lesssim 20$K 
yielding gaps of $\Delta=6$meV and $\Delta=20$meV, respectively.
Interestingly, however, an activation-law fit of the experimental resistivity fails above 20K.
We speculate that effects of the implanted helium atoms and the vicinity to the metal-insulator transition (dashed red line), in
combination with the very small gap, lead to a notable lifetime smearing of the spectral gap~\cite{PhysRevB.105.085139} already at temperatures $k_BT \ll\Delta$ and cause deviations from exponential behavior.
Additionally, we cannot exclude that the cfs-driven DMFT gap has an inherent temperature dependence due to the decreased inter-orbital fluctuations upon cooling.
Finally, vertex corrections to the conductivity may also play a non-trivial role and induce changes beyond the one-particle picture.

Samples with less than $3\cdot10^{15}$ He / cm$^2$ fluence ($c_\mathrm{exp} \le 4.00$\AA), instead exhibit Fermi liquid-like behavior, $\rho(T) \propto T^2$. Again, at low temperatures, small quantitative deviations occur that were interpreted as caused by disorder  renormalizations~\cite{PhysRevMaterials.3.115001}. Except for this additional low-temperature physics, the
 metallic regime is congruent with our DMFT phase diagram, left of the metal-insulator coexistence region, where the metal is stable against both magnetic and orbital order.
In this regime the variation of the fluence mainly affects the static disorder in the system, and thus the overall magnitude of the resistivity, not its overall temperature profile.

We end by displaying, in Fig.~\ref{fig:srvo3_cooling}, the evolution of the spectral functions for $c=4.05$\AA\ and $c=4.10$\AA\ from high (red) to low (blue) temperatures. This setting best describes the experimental situation---from synthesis at high $T$ to measurements at low $T$---of the metallic and insulating samples closest to the evidenced metal-insulator transition as a function of He-fluence~\cite{PhysRevMaterials.3.115001}.
In both cases, the high temperature solution ($T=770$K; red lines and points) is an incoherent insulator. Upon cooling (blue lines and dots), it transforms into either a paramagnetic metal ($c=4.05$\AA; left panel) or a coherent insulator ($c=4.10$\AA; right panel). We interpret the semi-quantitative agreement of the gap and temperature scales with experiment 
as supportive of our correlation-enhanced crystal-field scenario for the observed metal-insulator transition.

We note that crystal-field driven metal-insulator transitions have previously  been studied for symmetric two-band models in Ref.~\cite{PhysRevB.78.045115}, where a similar, direct transition from metal to orbitally polarized insulator was observed for small crystal-fields. Our setup hence generalizes these observation to the
case of t$_\mathrm{2g}$ orbitals that are more commonly encountered in correlated materials.
Indeed, on a material by material basis, the importance of crystal-field splittings onto the Mott transition has been highlighted before~\cite{keller:205116,pavarini:176403,poter_v2o3,pickemZoologySpinOrbital}. 
The He-implantation experiment provides a new means to manipulate the crystal-field in an almost continuous fashion.

\section{Conclusions}

Summarizing, our simulations mimicking He-irradiated SrVO$_3$ films on an SrTiO$_3$ substrate revealed the tetragonal distortion and the ensuing crystal-field splitting as the protagonist of the experimentally evidenced Mott transition.
Based on our realistic dynamical mean-field theory calculations, we established the complex $c$-axis parameter vs.\ temperature phase diagram shown in Fig.~\ref{fig:srvo3_sketch}. While it looks very similar to the iconic (DMFT) phase diagram of the one-orbital Hubbard model, the distinctly different control-parameter, the crystal-field splitting (or $c/a$-ratio), changes the nature of phases and phase boundaries: Instead of an anomaly in the overall charge compressibility, we propose that the crystal-field-driven redistribution of {\it partial} densities gives rise to an {\it orbital} Widom line. Further, the long-range order realized by super-exchange-like processes is not in the spin but in the charge channel, leading to orbital instead of magnetic order. Our work supports that neutral-ion implantation in correlated materials can be interpreted as chemical pressure, providing a pathway to study {\it expanded} structures complementary to the more accessible means to {\it compress} a material. 
Boundary conditions imposed by a substrate further enable an effectively negative  uniaxial pressure: a uniaxial expansion  which is otherwise hard to achieve.


\begin{acknowledgments}
 We thank 
J.~Gabel, R.~Claessen and M.~Sing
 for stimulating discussions surrounding the physics of oxide films.
 The authors acknowledge support from the Austrian Science Fund (FWF) through grants DOI 10.55776/P30213 and DOI 10.55776/I6142, SFB  Q-M\&S
 (grant DOI 10.55776/F86), and  research unit QUAST (grant DOI 10.55776/KIN2563725; for5249 of the German Research Foundation (DFG). Calculations were performed using the Austrian Scientific Computing (ASC) infrastructure.
 This work is largely based on Chapter 2.3.2 of MP's PhD thesis~\cite{Pickem_phd}.
\end{acknowledgments}

\appendix
\setcounter{figure}{0}

\renewcommand{\thefigure}{A\arabic{figure}}

\section{DFT band-structures} \label{app1}
\begin{figure*}[!t!h]
  \centering
  \includegraphics[width=0.8\linewidth]{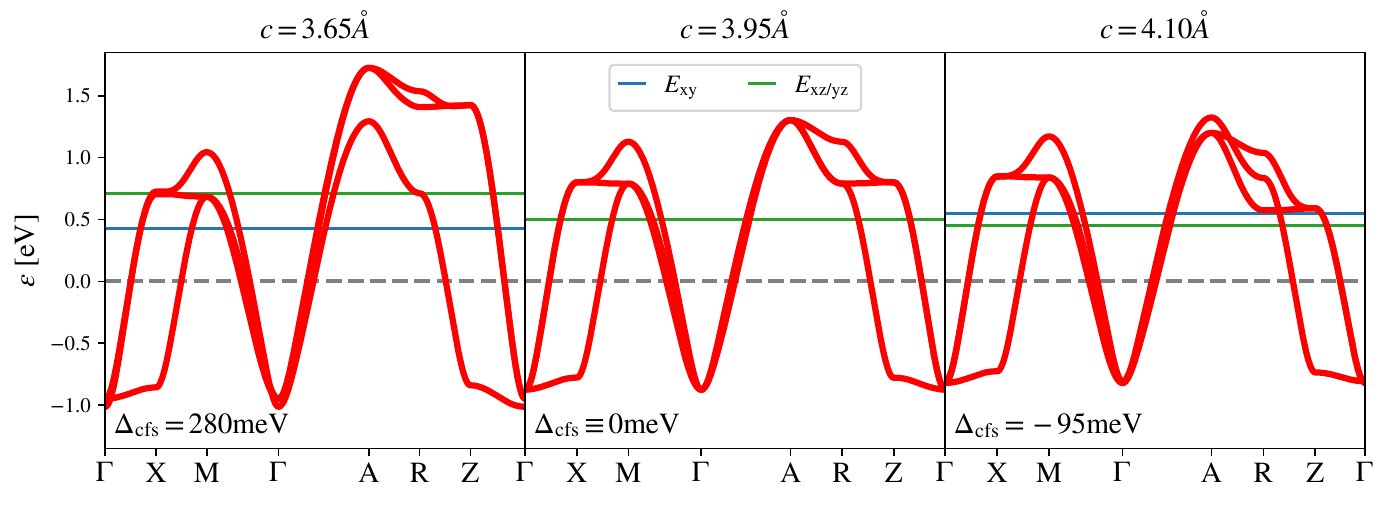}
  \caption{DFT band structure comparing $c=3.65$\AA~(left), $c=3.95$\AA~(middle) and $c=4.10$\AA~(right). Changes dominantly occur in the $k_z$ direction (k-path along $\Gamma$-A-R-Z-$\Gamma$) due to the changes to the out-of-plane orbital hybridization, affecting mostly {\it unoccupied} states. The local levels $E_\mathrm{xy}$ (blue) and $E_\mathrm{xz/yz}$ (green) are marked accordingly and showcase the reversal of the crystal-field splitting $\Delta_\mathrm{cfs} = E_{xz/yz}-E_{xy}$; the dashed grey line denotes the Fermi energy.}
 \label{fig:srvo3_dft_bscomp}
\end{figure*}

For selected examples, $c=3.65$\AA, $c=3.95$\AA, and $c=4.10$\AA~structures, 
we plot in Fig.~\ref{fig:srvo3_dft_bscomp} the DFT band-structure around the Fermi energy.
The dispersions within the in-plane range $\Gamma$-X-M-$\Gamma$ are hardly affected by the change in $c$-axis.
Still, looking more closely, the $\Gamma$-degeneracy is lifted, and the Fermi surface pocket surrounding the $\Gamma$-point is slightly modified. The structural changes are more apparent in the unoccupied states for $k_z > 0$: the $d_{xz}$/$d_{yz}$ orbital compression (elongation) widens (narrows) the dispersion by roughly $20$\% ($10$\%) for the shown cases with respect to the cubic case $c=3.95$\AA. 

\section{Supplementary DMFT spectra}
\begin{figure*}[!t!h]
  \centering
  \begin{subfigure}{0.49\textwidth}
  \raggedright (a) 
  \includegraphics[width=\linewidth]{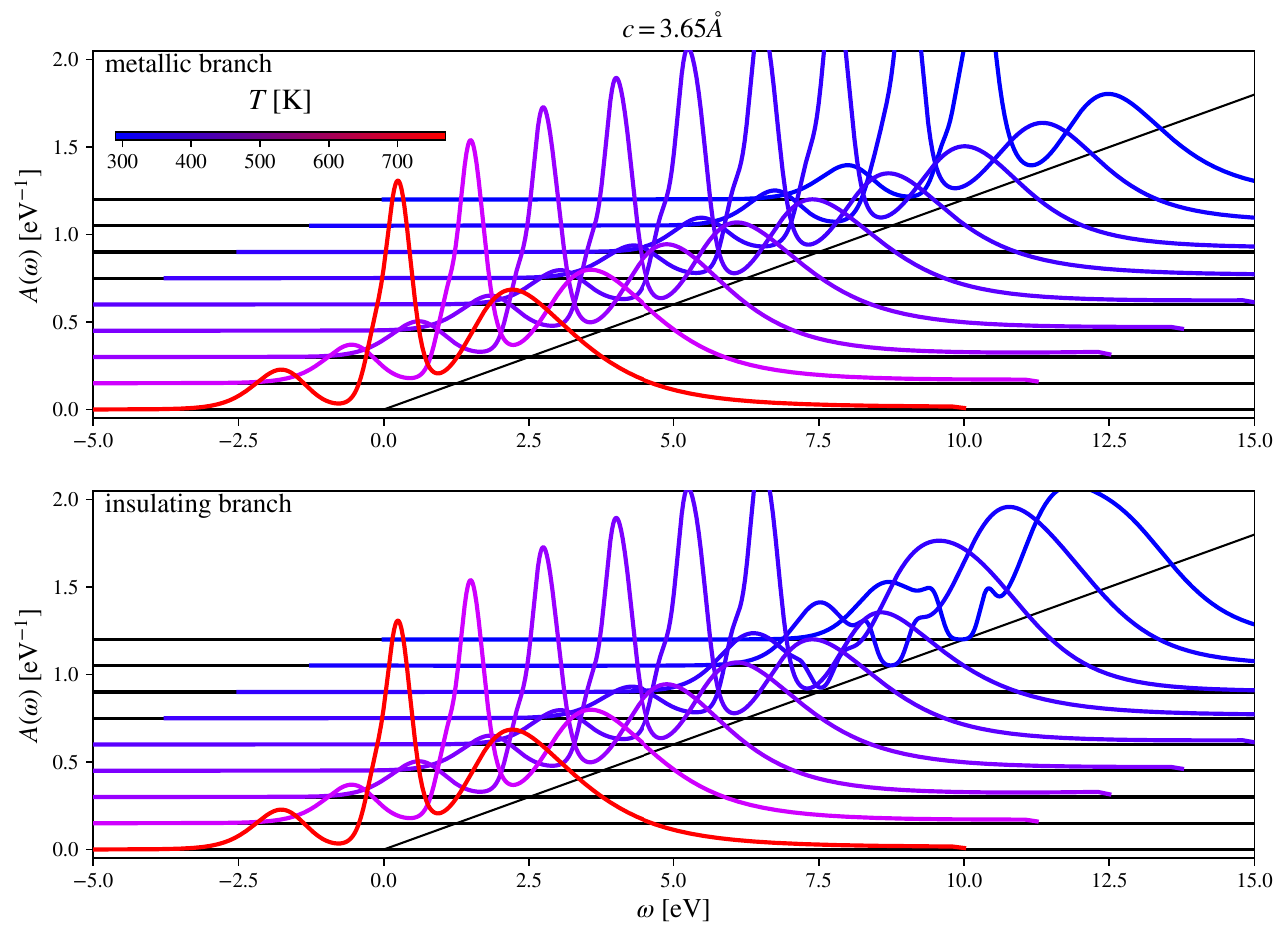}
  \phantomsubcaption\label{fig:srvo3_dmft_tempbranch1}
  \end{subfigure}
  \begin{subfigure}{0.49\textwidth}
  \raggedright (b) 
  \includegraphics[width=\linewidth]{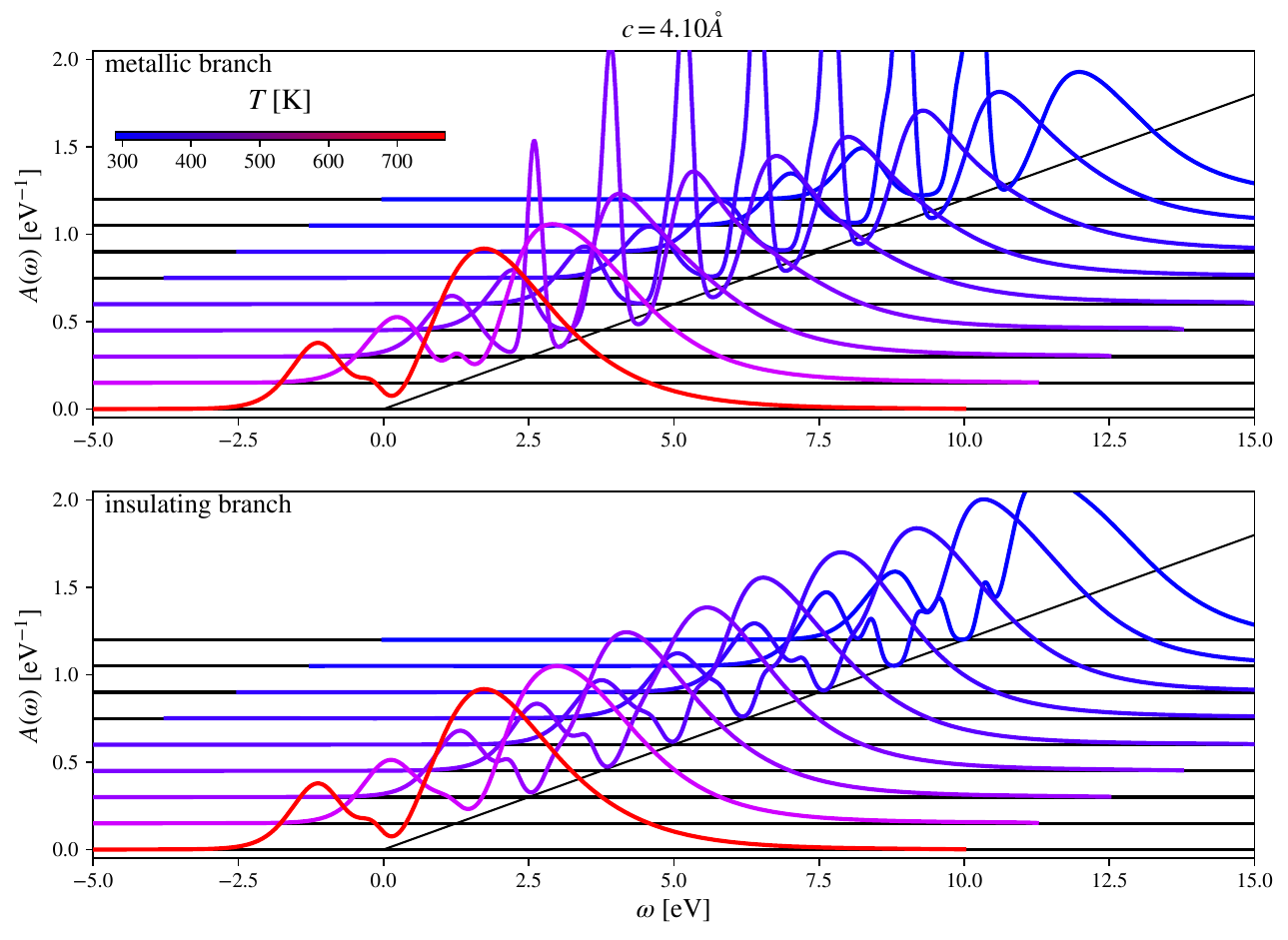}
  \phantomsubcaption\label{fig:srvo3_dmft_tempbranch2}
  \end{subfigure}
\caption{Left:
Spectral functions $A(\omega)$ for the two branches of the compressed structure $c=3.65$\AA~above $T=290$K, corresponding to the orbital occupations in the left panel of Fig.~\ref{fig:srvo3_temp}. Top: metallic branch. Bottom: insulating branch. The line color indicates the temperature (color map -- red: hot, blue: cold).
  Cooling the system from the high-temperature solution (red) the system maintains the well-formed three-peak structure on the metallic branch (as in the top panel). The onset of the insulating solution corresponds to the temperature at which an orbital polarization can be (forcefully) generated within DMFT.
  Right: Spectral function $A(\omega)$ for the two branches of the expanded structure $c=4.10$\AA~above $T=290$K, corresponding to the orbital occupations in the right panel of Fig.~\ref{fig:srvo3_temp}. Top: metallic branch. Bottom: insulating branch. Cooling the system from the high temperature ``bad insulator'' solution (red) leads to a coherent Mott insulator (insulating branch). Contrary to the compressed structure, the metallic solution is now the one that cannot be obtained via a temperature variation.
  \label{fig:highT}
  }
\end{figure*}

In this Appendix, we supplement our
temperature scans of Fig.~\ref{fig:srvo3_cooling} by considering in Fig.~\ref{fig:highT} quite extreme $c_{\mathrm{lattice}}$ parameters for which two solutions coexist. Hence we can
follow the spectral function on both metallic and insulating branches, separately.

\begin{figure}[!htb]
  \centering
\includegraphics[trim=4 0 4 0, clip=true, width=1.02\linewidth]{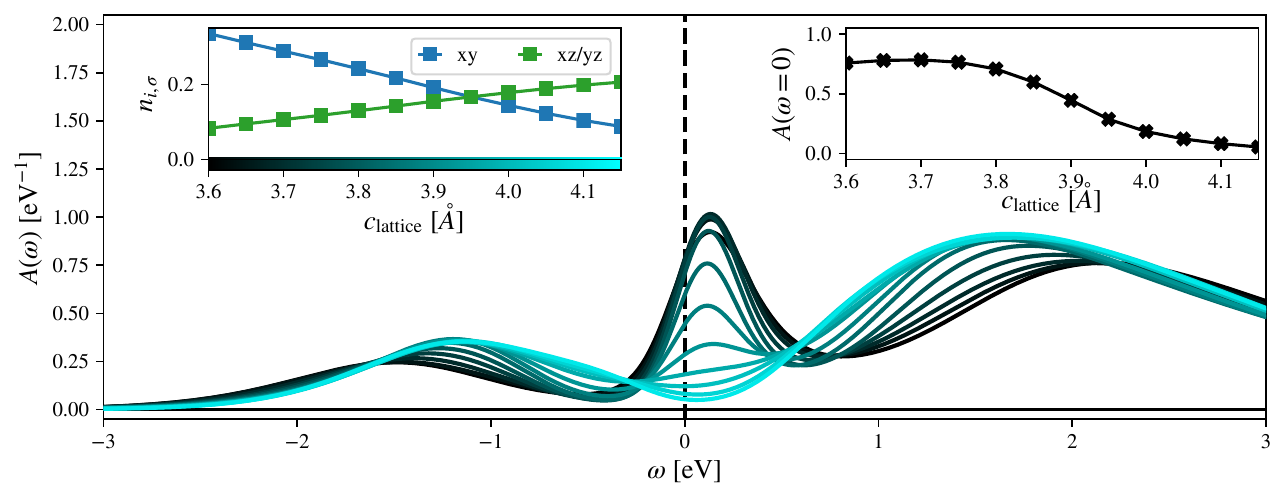}
  \caption{High temperature ($T=1160$K) spectral function scan as a function of $c_{\mathrm{lattice}}$ (see color bar of left inset). Left inset: orbital occupations. Right inset: spectral function at the Fermi level $A(\omega=0)$.
  The system undergoes a gradual crossover from metal ($c\leq 3.90$\AA) via bad metal ($3.90\AA \leq c \leq 4.0$\AA) to bad insulator $c\geq4.0$\AA.
 \label{fig:srvo3_hightemp2}
 }
\end{figure}

Further, in Fig.~\ref{fig:srvo3_hightemp2}
we show the spectral evolution at high temperatures, corresponding to 
 path {\textcircled{\raisebox{-.9pt} {1}}} in Fig.~\ref{fig:srvo3_sketch}.
At such a high temperature, no hysteresis but a smooth crossover from a metallic to the insulating solution is found upon increasing $c_{\mathrm{lattice}}$.
 
%

\end{document}